\documentclass[letterpaper, 10 pt, conference]{ieeeconf}  

\IEEEoverridecommandlockouts                              
\usepackage[utf8]{inputenc}
\usepackage[T1]{fontenc}
\usepackage{booktabs}

\usepackage[table]{xcolor}
\definecolor{rblue}{rgb}{0,0.5,1}
\definecolor{awesome}{rgb}{1.0, 0.13, 0.32}
\definecolor{hollywoodcerise}{rgb}{0.96, 0.0, 0.63}
\definecolor{lasallegreen}{rgb}{0.03, 0.47, 0.19}
\definecolor{hanpurple}{rgb}{0.32, 0.09, 0.98}
\definecolor{green(pigment)}{rgb}{0.0, 0.65, 0.31}

\makeatletter
\let\NAT@parse\undefined
\makeatother
\usepackage[pagebackref=false, breaklinks=true, colorlinks, bookmarks=false]{
        hyperref
}
\hypersetup{
        colorlinks=true,
        linkcolor={red},
        citecolor={hanpurple},
        urlcolor={magenta}
}

\usepackage{caption}
\usepackage{graphicx}
\usepackage{amsmath}
\usepackage{amssymb}
\usepackage{booktabs}
\usepackage{placeins}
\usepackage{stfloats}
\usepackage{url}

\title{\LARGE \bf
Beyond Isolated Entities: Relation-Aware Multi-Entity Modeling for Unsupervised Video Anomaly Detection
}

\author{Zhongpeng Pan$^{1}$, Xina Cheng$^{1,\dag}$, and Kailun Yang$^{2,3}$%
\thanks{This work was supported in part by the National Natural Science Foundation of China (Grant No. 62473139 and No. 62388101), in part by the Hunan Provincial Research and Development Project (Grant No. 2025QK3019), and in part by the State Key Laboratory of Autonomous Intelligent Unmanned Systems (the opening project number ZZKF2025-2-10).}
\thanks{$^{1}$The authors are with the School of Artificial Intelligence, Xidian University, China.}%
\thanks{$^{2}$The author is with the School of Artificial Intelligence and Robotics, Hunan University, China (email: kailun.yang@hnu.edu.cn).}%
\thanks{$^{3}$The author is also with the National Engineering Research Center of Robot Visual Perception and Control Technology, Hunan University, China.}%
\thanks{$^{\dag}$Corresponding author: Xina Cheng.}
}

\begin{document}

\maketitle
\thispagestyle{empty}
\pagestyle{empty}

\begin{abstract}
Video anomaly detection for patrol robots and surveillance systems must recognize abnormal interactions among familiar entities. Existing pixel-reconstruction and isolated-entity methods may fail when individual entities appear normal but their spatial or motion relations are abnormal. This work presents Interaction-Centric Network for Temporal Entity-Relation Analysis and Consistency Testing (INTERACT), a framework for unsupervised video anomaly detection. Person-Object Appearance-Motion Interaction (POAMI) moves beyond isolated-entity encoding by jointly modeling the appearance, motion, and spatial configurations of persons and objects in each frame. The resulting representations capture both individual entity states and their cross-entity context. Based on these representations, Target Geometry-Guided Relational Interaction Prediction (TGRIP) predicts target entity interaction states from historical relational memory and target-frame geometry without explicit identity tracking. Motion-Interaction Reconstruction and Alignment (MIRA) then evaluates these predictions through conditional flow reconstruction and semantic consistency checking, providing complementary anomaly evidence beyond prediction error alone. INTERACT achieves state-of-the-art performance, obtaining a frame-level AUC of $84.5\%$ on the ShanghaiTech benchmark. Ablation studies show that removing cross-entity attention causes the largest performance drop, demonstrating the necessity of relational modeling. INTERACT is particularly effective for anomalies caused by changes in relations among people and objects, while maintaining competitive performance in general scenarios. The source code will be made publicly available at \url{https://github.com/ppworkhard/INTERACT}. 
\end{abstract}

\section{Introduction}
Embodied robotic systems are increasingly deployed in patrol and surveillance settings---campus environments, industrial facilities, warehouses, and public spaces---where they must monitor dynamic scenes alongside humans, equipment, and everyday objects~\cite{cadena2016slam, villani2018survey}. Ensuring safe operation in such settings involves two complementary safety layers. \emph{Internal operational safety} concerns whether the robot's own actuators, sensors, and control pipeline are functioning correctly~\cite{schnell2022robigan, lasota2017safe}. \emph{External environmental safety}, the focus of this work, concerns whether the persons, objects, and their mutual interactions observed in the robot's visual stream conform to normal environmental patterns. Video anomaly detection is therefore essential for identifying abnormal events that may compromise safe operation in the observed environment.

Existing patrol systems can generally recognize predefined targets or flag known hazard categories. However, real-world patrol environments produce a long tail of abnormal events that cannot be exhaustively enumerated at deployment time---people falling, anomalous object handling, or sudden motion that disrupts historically stable configurations. Because anomalous events are rare and difficult to collect, video anomaly detection is commonly formulated as an unsupervised task that trains exclusively on normal video. The model must therefore infer normal patterns from normal observations while handling events that were not specified in advance. This setting creates three challenges.

\begin{figure}[!t]
    \centering
    \includegraphics[width=\columnwidth]{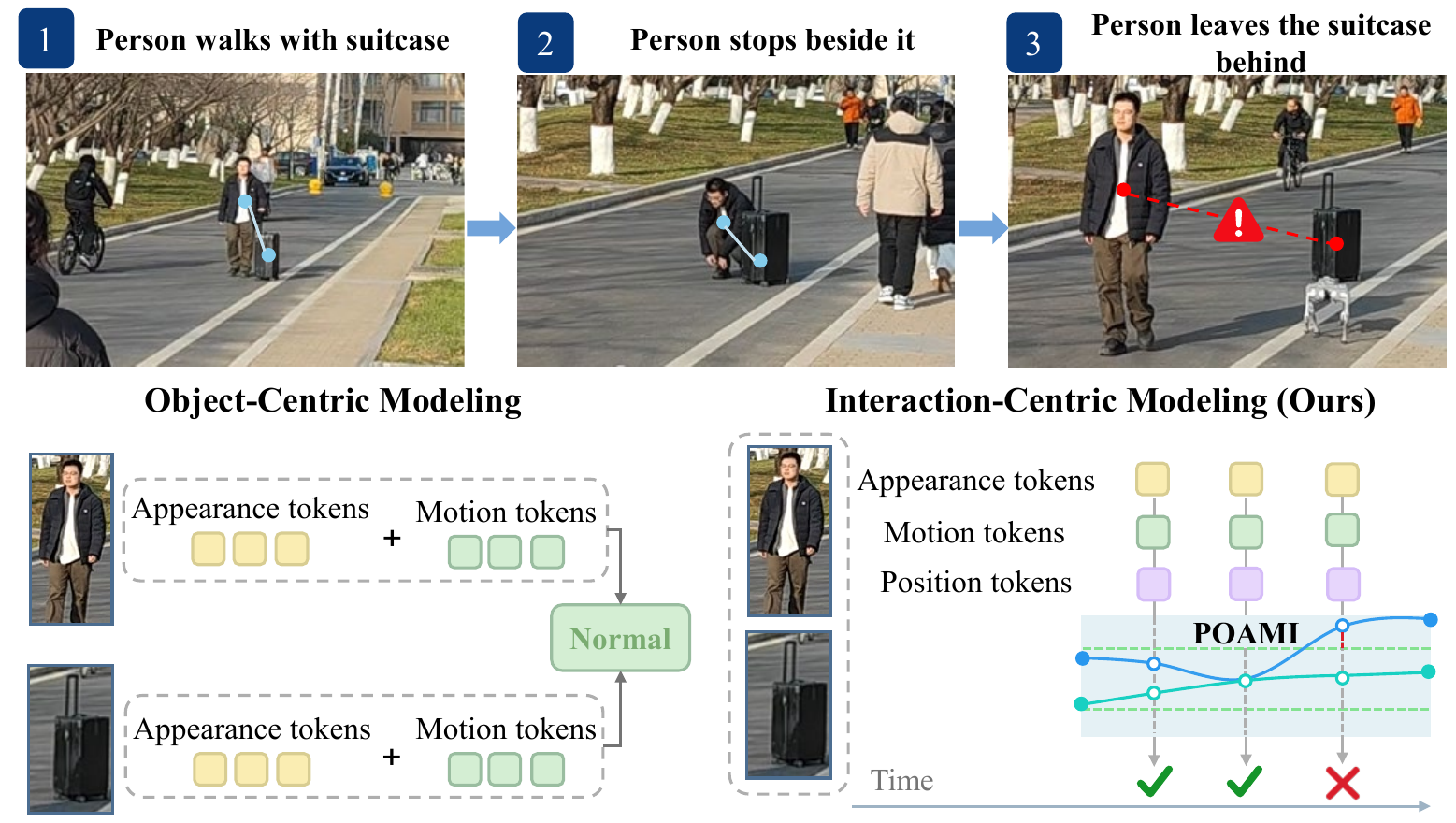}
    \caption{Conceptual illustration of relational anomaly detection. A person first walks with a suitcase, stops beside it, and then leaves it behind, forming an abnormal person--object relation although both entities remain individually familiar. Object-centric modeling (bottom left) processes entities and their appearance and motion cues independently, missing this relational change. INTERACT with POAMI (bottom right) jointly models the appearance, motion, and position of multiple entities; the curves show their evolving distances in the POAMI feature space, enabling detection of the abnormal interaction.}
    \label{fig:concept}
\end{figure}

Predicting the current interaction state from historical normal patterns is difficult when anomalies are open-ended and entity configurations change over time. Normal interactions vary across environments and entity combinations, while the number, positions, and participating entities may change between frames. Recent open-vocabulary detectors can recognize anomaly concepts beyond a predefined list~\cite{wuvad2024,anomize2025}. In the normal-only setting, however, the model must learn temporal regularities among entities and detect deviations from them. This calls for a prediction mechanism that accommodates changing entity sets; avoiding explicit identity association can further reduce dependence on consistent cross-frame matching.

Constructing relation-aware representations for heterogeneous entities poses a second challenge. A reliable detector must model persons and objects jointly, because anomalies can arise from changes in spatial arrangement, motion coherence, or temporal dependence even when individual entities appear normal. Capturing these relations requires integrating appearance, motion, position, and temporal context across entities; separately encoding these cues or naively concatenating them can miss cross-entity and cross-modal dependencies. Recent human-object interaction detectors model relational context and structured interaction clues~\cite{kim2023relctx,luo2024syntactic,jia2025contexthoi}, while recent video anomaly detectors combine complementary appearance and motion evidence~\cite{sdmae2024,twostream2024,ssae2025}. However, these approaches do not directly provide relation-aware representations for unsupervised detection of unseen relational deviations.

Verifying whether a predicted interaction is genuinely normal poses a third challenge. Reconstruction and future-prediction errors have distinct failure modes: reconstruction may be dominated by background texture. A prediction error only measures the difference between a forecast and the target. It does not show whether the forecast agrees with the entities' actual motion and interaction semantics. Reliable interaction-level scoring therefore requires complementary signals that assess both interaction predictability and motion-semantic consistency~\cite{twostream2024,adsm2025}.

Together, these challenges motivate an interaction-centric formulation of video anomaly detection. The formulation evaluates whether a multi-entity state is predictable from relational history and consistent with observed motion, while retaining sensitivity to abnormal entity appearance or motion. Figure~\ref{fig:concept} summarizes this distinction: individually normal entities can form an anomalous interaction when the relation between them changes. Normality is therefore modeled as \textbf{predictable and motion-consistent multi-entity interactions} rather than isolated entity appearances or motions.

\begin{figure*}[t]
    \centering
    \includegraphics[width=0.98\textwidth]{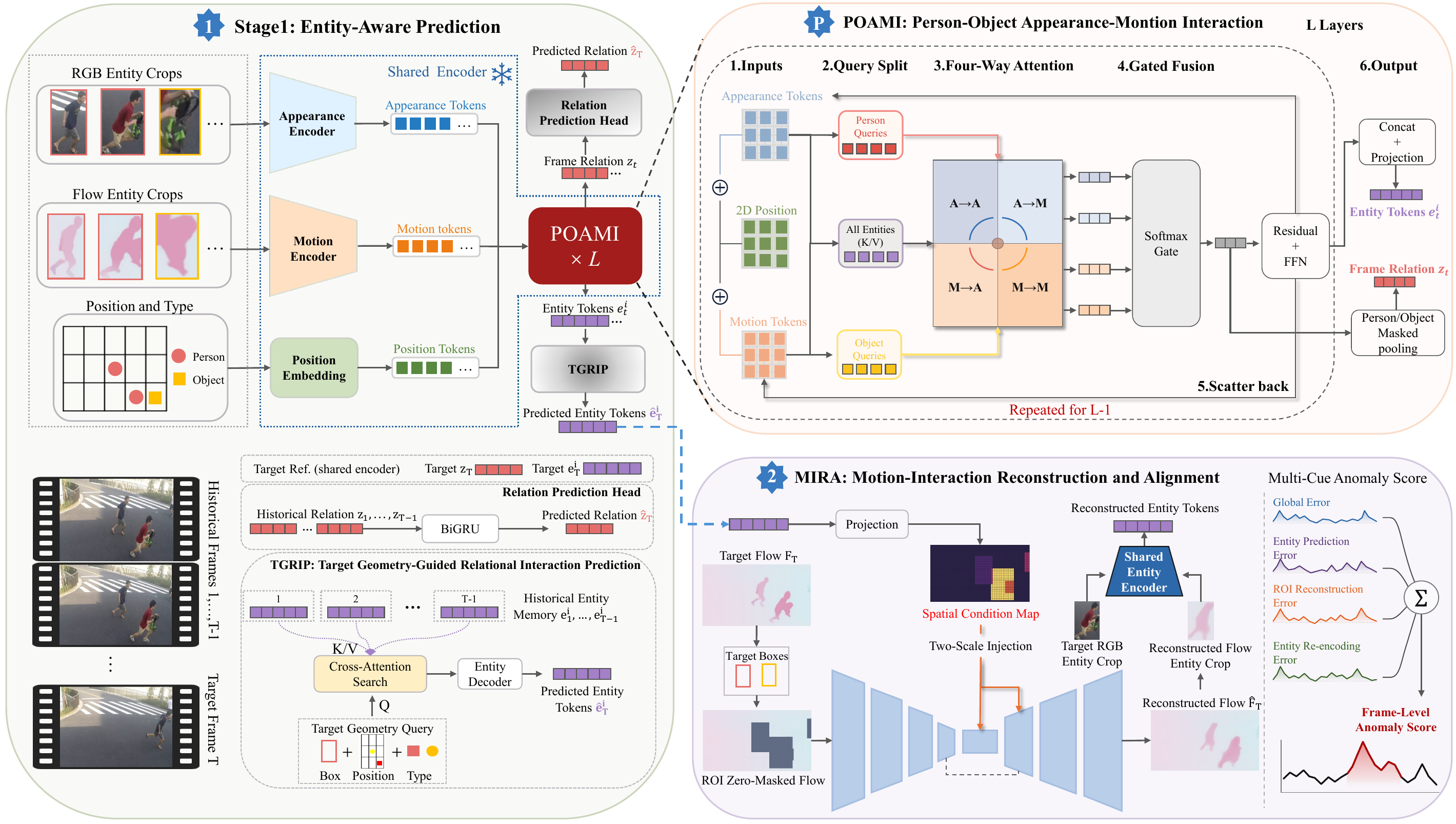}
    \caption{Overview of INTERACT. POAMI encodes multi-entity appearance, motion, and spatial relations. The Relation Prediction Head forecasts the global relation token, while TGRIP predicts entity tokens from relational history and target geometry. MIRA verifies the entity predictions through motion reconstruction and semantic alignment.}
    \label{fig:framework}
\end{figure*}

Building on this formulation, we develop Interaction-Centric Network for Temporal Entity-Relation Analysis and Consistency Testing (INTERACT), which assesses normality through the predictability of multi-entity interactions and their consistency with observed motion (Fig.~\ref{fig:framework}). To represent these interactions, we propose Person-Object Appearance-Motion Interaction (POAMI), which jointly encodes appearance, motion, and spatial relations across detected persons and objects. To model this temporal regularity under changing entity configurations, we design Target Geometry-Guided Relational Interaction Prediction (TGRIP). Conditioned on target-frame geometry, TGRIP predicts entity interaction representations from historical relational memory without explicit cross-frame identity tracking. To assess whether these predictions explain the observed motion, we introduce Motion-Interaction Reconstruction and Alignment (MIRA). MIRA conditions motion reconstruction on the predicted representations and assesses semantic consistency in the learned relation-aware representation space. Together, INTERACT combines deviations from expected interaction states with motion reconstruction and representation consistency errors to identify departures from learned normal patterns.

Extensive experiments on ShanghaiTech, Avenue, and NWPU Campus\_sub, a manually selected subset of NWPU Campus, demonstrate the effectiveness of INTERACT. INTERACT achieves state-of-the-art frame-level AUC on ShanghaiTech with \(84.5\%\), while maintaining competitive performance on the other datasets. Ablation studies confirm the necessity of relational modeling and the complementary benefits of prediction and verification. A preliminary evaluation on first-person videos captured by a Unitree Go2 quadruped robot without retraining further demonstrates INTERACT's potential for mobile robotic monitoring.

In summary, our key contributions are:
\begin{itemize}
    \item We propose a relation-centric anomaly detection paradigm that extends normality modeling from isolated entity regularity to multi-entity interaction regularity, jointly defined by interaction predictability and motion interpretability.
    \item We introduce POAMI, a relation-aware multi-entity representation that jointly models appearance, motion, and spatial position through cross-modal attention, producing entity tokens that encode both entity state and inter-entity context.
    \item We present TGRIP, which handles dynamic entity sets without explicit identity tracking, and MIRA, which grounds predicted interactions in physical motion and relational semantics.
    \item Extensive experiments on ShanghaiTech~\cite{luo2017revisit} demonstrate state-of-the-art frame-level AUC of \(84.5\%\), with ablations confirming the necessity of each component.
\end{itemize}

\section{Related Work}

\subsection{Robot Anomaly and Safety Monitoring}

Safety monitoring for autonomous platforms has developed primarily around the robot's own internal state. Model- and signal-based fault diagnosis methods track proprioceptive streams---joint torques, encoder residuals, and control-loop errors---and declare a fault when observed signals depart from an analytical or learned model of nominal operation~\cite{khalastchi2018fault}. Learning-based variants replace hand-designed residuals with generative models of normal telemetry: recurrent variational autoencoders score multimodal sensor sequences during manipulation~\cite{park2018multimodal}, and bidirectional adversarial models measure reconstruction discrepancy in latent space~\cite{schnell2022robigan}. A parallel line in human-robot collaboration enforces external safety through geometric and kinematic constraints, using proximity, speed-and-separation, and predicted human motion to modulate robot behaviour~\cite{lasota2017safe, villani2018survey}. Both families are effective within their scope, but their notion of normality is tied either to the robot's own actuation signals or to a metric safety envelope around the human. Neither judges whether the \emph{semantic} configuration of persons and objects in the observed scene is consistent with how that environment normally behaves, which is what a patrol robot must assess.

\subsection{Video Anomaly Detection}

Unsupervised video anomaly detection learns normality from anomaly-free video and scores deviation at test time. Reconstruction-based methods train autoencoders on normal frames or cuboids and treat reconstruction error as the anomaly signal~\cite{hasan2016learning,sdmae2024}. Memory-augmented models constrain the latent space with normal prototypes~\cite{gong2019memorizing,park2020learning}, while recent methods use scene-dependent latent prediction and denoising score matching~\cite{ssae2025,adsm2025}. Prediction-based methods learn temporal regularity by forecasting a future frame under appearance, gradient, and optical-flow constraints~\cite{liu2018future}; hybrid designs couple flow reconstruction with flow-guided frame prediction~\cite{liu2021hybrid,twostream2024}.
A third group replaces pixel residuals with self-supervised pretext tasks, including spatio-temporal jigsaw puzzles and multi-task proxy objectives over motion and appearance~\cite{wang2022video,georgescu2021anomaly}. Weakly supervised formulations relax the normal-only assumption using video-level labels~\cite{sultani2018real,tian2021weakly}, at the cost of anomalous footage that patrol deployments cannot enumerate in advance.
Across these mechanisms the scored quantity is a frame- or region-level residual; even when the encoder observes several entities, error is aggregated over pixels rather than attributed to a relation between entities, so a divergence between two individually plausible trajectories leaves little residual to detect.

\subsection{Entity-Centric and Interaction Modeling}

To suppress background bias, object-centric methods restrict normality modeling to detected regions. Object-level autoencoders with appearance and motion branches score instances independently~\cite{ionescu2019object}, and background-agnostic frameworks add adversarial training and object-specific reconstruction targets~\cite{georgescu2022background}. More recent skeleton- and pose-based methods use motion-conditioned diffusion to model normal dynamics~\cite{mocodad2023}, while scene-aware contrastive formulations relate entity semantics to their surrounding context~\cite{sun2023hierarchical}. These approaches do encode relational and contextual structure---graph edges over joints, scene-conditioned semantics---but the relations are largely intra-entity or entity-to-background, and the objective remains reconstruction or clustering of individual instances. In parallel, recent human-object interaction detectors model relational context and structured interaction clues~\cite{kim2023relctx,luo2024syntactic,jia2025contexthoi}. Such models are relation-aware by construction, yet they are supervised classifiers over a closed interaction vocabulary, so they can only name interactions that were annotated during training and offer no mechanism for judging an unseen configuration as abnormal.

INTERACT sits between these lines. We keep the unsupervised, normal-video-only setting, but move the modelled unit from the frame or isolated instance to the multi-entity relation: a relation-aware representation jointly encodes appearance, motion, and spatial position over the full entity set, normality is defined by whether the current interaction state is predictable from relational history, and predicted interactions must further be consistent with observed motion and relational semantics~\cite{yang2026tau}. Anomaly evidence therefore comes from violated relations among familiar entities rather than from pixel residuals or membership in a supervised interaction class.

\section{Methodology}

\subsection{Problem Formulation and INTERACT Overview}

We address unsupervised video anomaly detection in patrol settings where only normal training video is available. The input is a video clip of length $T$. Frame $t$ contains an RGB image $I_t$, three-channel optical flow $F_t$, and entity bounding boxes $b_{t,i}$ produced independently by a pretrained YOLOv8 detector. Each detection yields an RGB patch $R_{t,i}$, a flow patch $F_{t,i}$, normalized box geometry, and a person-or-object class. The interaction state at frame $t$ is represented by a set of entity tokens $\{z_{t,i}\}$ and a global relation token $z_t^g$. Our goal is to predict the target state at frame $T$ from the preceding $T-1$ frames and use its deviation as anomaly evidence. INTERACT does not perform explicit cross-frame identity tracking.

INTERACT is optimized in two stages. Stage~1 predicts global and entity interaction states, while Stage~2 uses MIRA to verify entity predictions with Stage~1 frozen.

\subsection{POAMI: Relation-Aware Multi-Entity Appearance-Motion Representation}
\label{sec:poami}

To capture interaction normality, each entity representation must encode not only its own appearance and motion, but also how these cues relate across the entity set. POAMI provides this relation-aware encoding by jointly modeling appearance, motion, and spatial structure.

Two convolutional encoders extract appearance and motion features independently:
\begin{equation}
x^g_{t,i} = E_g(R_{t,i}), \quad x^m_{t,i} = E_m(F_{t,i}).
\end{equation}
Each bounding-box center is discretized on a $30\times54$ spatial grid. A fixed two-dimensional sine-cosine encoding, indexed by the resulting grid coordinate, is added to both $x^g_{t,i}$ and $x^m_{t,i}$. We divide valid detections into person entities $\mathcal{P}_t$ and object entities $\mathcal{O}_t$. POAMI then computes two intra-modal streams, appearance-to-appearance $A_{GG}$ and motion-to-motion $A_{MM}$, and two cross-modal streams, appearance-to-motion $A_{GM}$ and motion-to-appearance $A_{MG}$. Both person and object queries attend over the complete entity set $\mathcal{P}_t\cup\mathcal{O}_t$, allowing every entity to gather context across classes. A learned softmax gate fuses the four streams. The fused response is projected back to the appearance and motion branches through residual connections, followed by separate feed-forward updates. The final entity token $z_{t,i}$ is obtained by concatenating the updated features and projecting them to $d_{\text{entity}}=1024$. A global relation token $z_t^g$ is formed by separately pooling person and object relation features and fusing them with $w_{\text{obj}}=0.2$. When no person is detected, the object representation is used as the global token. POAMI therefore encodes individual state and multi-entity context without explicit graph construction or identity association.

\begin{figure}[!t]
    \centering
    \includegraphics[width=\columnwidth]{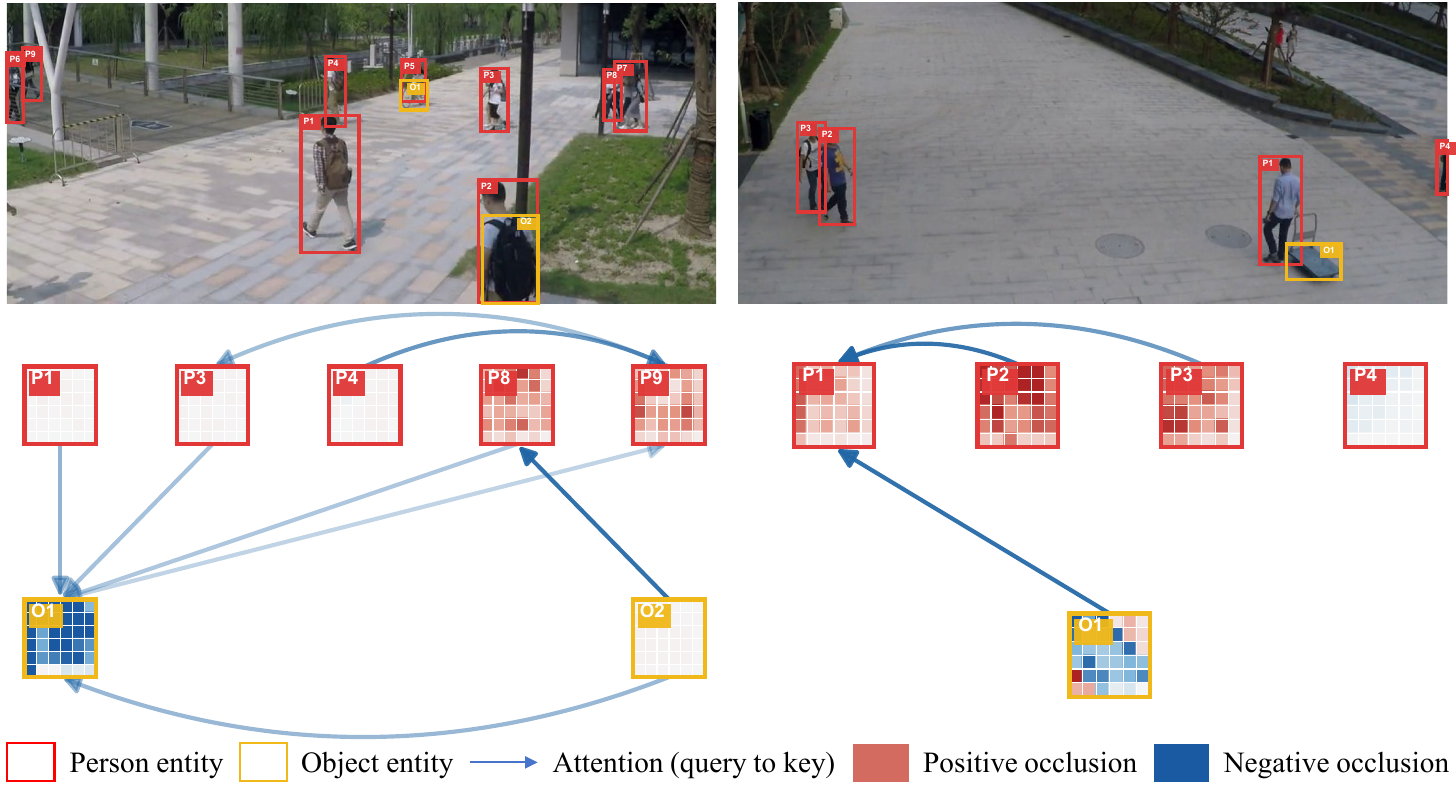}
    \caption{Cross-entity attention and occlusion sensitivity in scenes with more people (left) and fewer people (right). Red/yellow boxes denote persons/objects, and arrows show stronger attention connections retained for readability. Heatmap colors indicate increases (red), decreases (blue), or minimal changes (near-white) in the Stage~1 anomaly score after local occlusion.}
    \label{fig:entity_relations}
\end{figure}

Fig.~\ref{fig:entity_relations} illustrates the resulting cross-entity attention in scenes with different numbers of people. The displayed person--person and person--object connections show how entities aggregate contextual information from the surrounding entity set.

\subsection{Dual-Branch Relational Interaction Prediction}
\label{sec:prediction}

Normal interactions exhibit temporal regularity, so the target configuration should be predictable from relational history. Stage~1 models this predictability with separate global and entity branches.

\textbf{Relation Prediction Head.} Historical global tokens $\{z^g_1,\ldots,z^g_{T-1}\}$ are processed by a bidirectional GRU to predict $\hat{z}_T^g$. The dimension-normalized prediction loss is
\begin{equation}
\mathcal{L}_{\text{global}}=\frac{1}{d_g}\|\hat{z}_T^g-z_T^g\|_2^2,
\end{equation}
where $d_g$ is the global-token dimension. This branch supplies a scene-level relation prediction signal.

\textbf{TGRIP.} Historical entity tokens $\{z_{t,i}\}_{t<T}$ are concatenated with temporal embeddings to form relational memory. For each target entity $j$, its discretized center coordinate is normalized by the spatial-grid dimensions. The normalized center, box extent, and person-or-object class form a geometry query. Query self-attention captures the target configuration, and cross-attention retrieves relevant historical relations from memory to produce $\hat{z}_{T,j}$. This design accommodates changing entity counts and positions without explicit ID matching. The entity-token prediction loss is
\begin{equation}
\mathcal{L}_{\text{entity}} = \frac{1}{|\mathcal{V}_T|d_{\text{entity}}}\sum_{j \in \mathcal{V}_T} \|\hat{z}_{T,j} - z_{T,j}\|_2^2,
\end{equation}
where $\mathcal{V}_T$ denotes the valid target-frame entities. An exponential-moving-average (EMA) teacher provides stable target entity tokens during Stage~1 training.

The Stage~1 training objective combines four loss terms:
\begin{equation}
\mathcal{L}_{\text{S1}} = \lambda_g \mathcal{L}_{\text{global}} + \lambda_e \mathcal{L}_{\text{entity}} + \lambda_v \mathcal{L}_{\text{var}} + \lambda_c \mathcal{L}_{\text{cov}},
\end{equation}
where $\mathcal{L}_{\text{var}}$ and $\mathcal{L}_{\text{cov}}$ regularize the global relation sequence to prevent representation collapse. We use $\lambda_g=1.0$, $\lambda_e=0.5$, $\lambda_v=1.0$, and $\lambda_c=0.001$. The EMA teacher uses decay $\tau=0.99$; during training, its target-frame encoding replaces $z_{T,j}$ in $\mathcal{L}_{\text{entity}}$.

\subsection{MIRA: Motion-Interaction Reconstruction and Alignment}
\label{sec:verification}

Prediction error alone cannot confirm whether a forecast is physically grounded. MIRA checks consistency in both motion and semantic spaces.

\textbf{Motion-space verification.} We zero the target-frame flow within entity ROIs and reconstruct the full flow conditioned on predicted entity tokens $\{\hat{z}_{T,j}\}$. Each token is projected to $d_{\text{cond}}=128$ and rasterized within its bounding box to form a spatial condition map. A conditional flow autoencoder takes the masked flow and condition map as input and produces $\hat{F}_T$. During training, predicted tokens are shuffled across the batch to construct incorrect conditions. A margin loss requires the correct conditions to yield lower ROI error, encouraging sensitivity to interaction information. The shuffled branch is used only for this training constraint and does not produce an additional inference score.

\textbf{Semantic-space verification.} Target RGB patches and reconstructed flow patches are re-encoded by the frozen POAMI module to obtain $\tilde{z}_{T,j}$. These tokens are compared with $z_{T,j}$ encoded from the observed target RGB and flow. Thus, semantic consistency is defined by POAMI representations rather than manual interaction labels.

The Stage~2 training objective is:
\begin{equation}
\mathcal{L}_{\text{S2}} = \lambda_{\text{roi}} \mathcal{L}_{\text{roi}} + \lambda_{\text{img}} \mathcal{L}_{\text{img}} + \lambda_{\text{margin}} \mathcal{L}_{\text{margin}} + \lambda_{\text{cons}} \mathcal{L}_{\text{cons}},
\end{equation}
Let $M_T$ be the union of valid entity masks and $C$ the number of flow channels. The ROI and semantic consistency losses are normalized over valid elements:
\begin{equation}
\begin{aligned}
\mathcal{L}_{\text{roi}}&=\frac{\|(\hat{F}_T-F_T)\odot M_T\|_1}{C\|M_T\|_1},\\
\mathcal{L}_{\text{cons}}&=\frac{1}{|\mathcal{V}_T|d_{\text{entity}}}
\sum_{j\in\mathcal{V}_T}\|\tilde{z}_{T,j}-z_{T,j}\|_2^2.
\end{aligned}
\end{equation}
The full-image loss is $\mathcal{L}_{\text{img}}=\operatorname{mean}|\hat{F}_T-F_T|$. The condition-sensitivity loss is $\mathcal{L}_{\text{margin}}=\max(0,m+\mathcal{L}_{\text{roi}}^{\text{correct}}-\mathcal{L}_{\text{roi}}^{\text{shuffled}})$. We use $\lambda_{\text{roi}}=1.0$, $\lambda_{\text{img}}=0.1$, $\lambda_{\text{margin}}=0.2$, $\lambda_{\text{cons}}=0.05$, and $m=0.07$. Stage~1 is frozen, and only the conditional autoencoder is updated.

\subsection{Anomaly Score Fusion and Inference}
\label{sec:score}

At test time, we first compute dimension-normalized errors for each valid target entity:
\begin{equation}
e^{\text{pred}}_j=\frac{\|\hat{z}_{T,j}-z_{T,j}\|_2^2}{d_{\text{entity}}}, \quad
e^{\text{cons}}_j=\frac{\|\tilde{z}_{T,j}-z_{T,j}\|_2^2}{d_{\text{entity}}},
\end{equation}
\begin{equation}
e^{\text{roi}}_j=\frac{\|(\hat{F}_T-F_T)\odot M_{T,j}\|_1}{C\|M_{T,j}\|_1}.
\end{equation}
The entity-level signals are aggregated by their maximum, while the global signal uses the global-token prediction error:
\begin{equation}
\begin{aligned}
s_{\text{entity}}&=\max_{j\in\mathcal{V}_T}e^{\text{pred}}_j, &
s_{\text{roi}}&=\max_{j\in\mathcal{V}_T}e^{\text{roi}}_j,\\
s_{\text{entity\_rec}}&=\max_{j\in\mathcal{V}_T}e^{\text{cons}}_j, &
s_{\text{global}}&=\frac{\|\hat{z}_T^g-z_T^g\|_2^2}{d_g}.
\end{aligned}
\end{equation}
These four signals are combined via fixed manually-set weights:
\begin{equation}
S_t = w_e\, s_{\text{entity}} + w_g\, s_{\text{global}} + w_r\, s_{\text{roi}} + w_c\, s_{\text{entity\_rec}}.
\end{equation}
We use $w_e=0.30$, $w_g=0.10$, $w_r=0.10$, $w_c=0.50$. The weights reflect that entity relation prediction and semantic reconstruction capture primary interaction deviations, while global relation and motion reconstruction provide auxiliary verification. No per-signal normalization (\textit{e.g.}, min-max or z-score) is applied. For overlapping clips mapping to the same frame, scores are averaged by coverage count. Final frame-level scores undergo temporal smoothing with a Gaussian filter (sigma=8, mode=nearest, truncate=4) before evaluation.

Together, POAMI, the Relation Prediction Head, and TGRIP model and forecast interaction states at entity and global levels. MIRA then tests the entity forecasts against observed motion and POAMI semantics. Their four residual signals yield a frame-level score for deviations from learned interaction patterns.

\begin{table*}[!t]
\centering
\caption{Frame-level ROC-AUC (\%) comparison with existing methods on video anomaly detection benchmarks. INTERACT achieves state-of-the-art performance on ShanghaiTech~\cite{luo2017revisit} ($84.5\%$) while maintaining competitive results on Avenue~\cite{lu2013abnormal} and NWPU Campus\_sub~\cite{wang2023nwpu}.}
\label{tab:main_comparison}
\begin{tabular}{lcccc}
\toprule
Method & Venue & Avenue & ShanghaiTech & NWPU Campus \\
\midrule
FFP~\cite{liu2018future} & CVPR18 & 84.9 & 72.8 & -- \\
MemAE~\cite{gong2019memorizing} & ICCV19 & 84.9 & 72.8 & 61.9 \\
CAE~\cite{ionescu2019object} & CVPR19 & 87.4 & 78.7 & -- \\
MNAD~\cite{park2020learning} & CVPR20 & 88.5 & 70.5 & 62.5 \\
AMMC-Net~\cite{ammcnet2021} & AAAI21 & 86.6 & 73.7 & 64.5 \\
HF$^{2}$-VAD~\cite{liu2021hybrid} & ICCV21 & 91.1 & 76.2 & 63.7 \\
VABD~\cite{vabd2022} & TIP22 & 86.6 & 78.2 & -- \\
MPN~\cite{lv2021learning} & CVPR21 & 89.5 & 73.8 & 64.4 \\
SSMTL~\cite{georgescu2021anomaly} & CVPR21 & 91.5 & 82.4 & -- \\
MoCoDAD~\cite{mocodad2023} & ICCV23 & 89.0 & 77.9 & -- \\
FBAE~\cite{wang2023nwpu} & CVPR23 & 86.8 & 79.2 & 68.2 \\
FPDM~\cite{fpdm2023} & ICCV23 & 90.1 & 78.6 & -- \\
HSC~\cite{sun2023hierarchical} & CVPR23 & \textbf{92.4} & 83.4 & -- \\
SLMPT~\cite{slmpt2023} & ICCV23 & 90.9 & 78.8 & -- \\
Two-stream~\cite{twostream2024} & TIP24 & 90.8 & 83.7 & -- \\
SDMAE~\cite{sdmae2024} & CVPR24 & 91.3 & 79.1 & -- \\
SSAE~\cite{ssae2025} & TPAMI25 & 90.2 & 80.5 & \underline{75.6} \\
MA-PDM~\cite{mapdm2025} & AAAI25 & 91.3 & 79.2 & -- \\
Noise-Resistant VAD~\cite{noisevad2025} & CVPR25 & \underline{92.1} & \underline{84.3} & -- \\
ADSM~\cite{adsm2025} & ICCV25 & 91.6 & \textbf{84.5} & \textbf{76.9} \\
\midrule
INTERACT & -- & 90.6 & \textbf{84.5} & 72.4$^{\dag}$ \\
\bottomrule
\end{tabular}
\vspace{2pt}
\parbox[t]{0.72\textwidth}{\footnotesize Bold and underlined values indicate the best and second-best results, respectively. $^{\dag}$Our result is evaluated on NWPU Campus\_sub, whereas the reported results of existing methods use the original NWPU Campus test set; these results are not directly comparable.}
\end{table*}

\section{Experiments}

\subsection{Datasets}

We evaluate INTERACT on ShanghaiTech, Avenue, and NWPU Campus\_sub, where the last is a manually curated evaluation subset derived from the NWPU Campus test set.

\textbf{ShanghaiTech}~\cite{luo2017revisit} is a large-scale campus surveillance dataset comprising $330$ training videos and $107$ test videos captured from $13$ fixed camera views. Training videos contain only normal events; test videos include $130$ annotated anomalous events spanning fighting, chasing, vehicle intrusion, and unusual pedestrian behavior. We used the official train-test split and evaluated frame-level anomaly detection performance.

\textbf{Avenue}~\cite{lu2013abnormal} consists of $16$ training and $21$ test video clips recorded in a pedestrian walkway. Training clips depict typical walking patterns; test clips include anomalies such as throwing objects, running, and wrong-direction motion. We followed the standard protocol, training exclusively on normal frames and reporting frame-level AUC on the full test set.

\textbf{NWPU Campus\_sub}~\cite{wang2023nwpu} is a manually curated evaluation subset derived from the NWPU Campus test set. The NWPU Campus dataset was introduced by Wang et al.~\cite{wang2023nwpu} as a large-scale video anomaly detection benchmark featuring diverse indoor and outdoor campus scenes, multiple camera viewpoints, lighting conditions, and event contexts across different times and locations. Because NWPU Campus contains many scene-context-dependent anomalies where abnormality is tied to specific locations or environmental cues, we manually selected a subset to reduce this bias. The selection criterion focuses on scene-independent anomalies that primarily involve abnormal entity behaviors or interactions, such as climbing, stealing, fighting, chasing, and crowd gathering. We retain $118$ normal videos and select $64$ anomalous videos that meet this criterion. The resulting subset contains $182$ test videos and is used as the third evaluation set in this work.

\subsection{Evaluation Protocol}
Each clip comprised $16$ frames: the first $15$ as historical context and frame $16$ as the prediction target.
Training and test clips use strides of 8 and 1, respectively. Frame scores follow the fusion and smoothing procedure in Section~\ref{sec:score}. Frame-level ROC-AUC pools all test frames within each evaluation set.
\subsection{Implementation Details}

We use precomputed YOLOv8 detections and RAFT optical flow. Bounding boxes are filtered with a confidence threshold of $0.5$; class 0 is designated as person and all other classes as objects. Entities are ordered with persons first, followed by objects. Flow patches contain horizontal motion ($u$), vertical motion ($v$), and magnitude $\sqrt{u^2+v^2}$. RGB and flow patches are cropped according to the bounding boxes, resized to $96\times96$, and RGB values are normalized to $[0,1]$. The spatial grid for position encoding is determined by dividing the frame resolution into $16\times16$ patches, resulting in $30\times54$ for ShanghaiTech ($856\times480$ frames).

The appearance and motion encoders are independent CNNs. 
Relation features are $512$-dimensional, and entity tokens are $1024$-dimensional.
The relation stack uses $3$ layers with $4$ attention heads, and the entity predictor uses $2$ layers with $8$ heads. For Stage~2, the conditional flow autoencoder receives full-frame flow at $256\times448$ resolution and applies zero-mask degradation within detected bounding boxes. Stage~1 is trained with Adam using batch size $12$ and learning rate $10^{-4}$; Stage~2 freezes Stage~1 and uses Adam with batch size $16$ and learning rate $2\times10^{-5}$. All experiments use a single NVIDIA GeForce RTX 4090.

\subsection{Comparison with Existing Methods}

Table~\ref{tab:main_comparison} compares INTERACT with existing approaches. 
On ShanghaiTech, INTERACT achieved $84.5\%$ AUC, matching the best reported result. Many anomalies in ShanghaiTech involve multi-person interactions or coordinated person-object events such as chasing, vehicle intrusion, and cycling, which is consistent with our modeling bias toward relational patterns.
On Avenue, INTERACT achieved $90.6\%$ AUC, which is competitive but not the highest. Many Avenue anomalies primarily involve individual motion patterns such as running and jumping, which align less directly with our emphasis on cross-entity interactions.
On NWPU Campus\_sub, INTERACT achieved $72.4\%$ AUC.
The manually curated subset focuses on scene-independent interaction anomalies while reducing location-specific cues. The result demonstrates that INTERACT maintains reasonable detection capability across diverse anomaly types, even when the subset composition differs from the specific interaction patterns emphasized in our modeling.

\subsection{Ablation Studies}

\begin{table*}[!t]
\centering
\caption{Ablation study on ShanghaiTech. Delta is computed relative to the full INTERACT model (A0).}
\label{tab:ablation}
\begin{tabular}{clcccccc}
\toprule
ID & Variant & \begin{tabular}{@{}c@{}}Cross-entity\\Attn.\end{tabular} & \begin{tabular}{@{}c@{}}Prediction\\Score\end{tabular} & Verification & \begin{tabular}{@{}c@{}}Token\\Consistency\end{tabular} & AUC (\%) & $\Delta$ \\
\midrule
A0 & INTERACT (Full) & \checkmark & \checkmark & \checkmark & \checkmark & 84.50 & 0.00 \\
A1 & Entity-Independent Variant & $\times$ & \checkmark & \checkmark & \checkmark & 79.18 & -5.32 \\
A2 & Prediction Only & \checkmark & \checkmark & $\times$ & -- & 80.93 & -3.57 \\
A3 & Verification Only & \checkmark & $\times$ & \checkmark & \checkmark & 82.93 & -1.57 \\
A4 & Without Token Consistency & \checkmark & \checkmark & \checkmark & $\times$ & 83.43 & -1.07 \\
\bottomrule
\end{tabular}
\end{table*}

\begin{figure*}[!t]
    \centering
    \includegraphics[width=\textwidth]{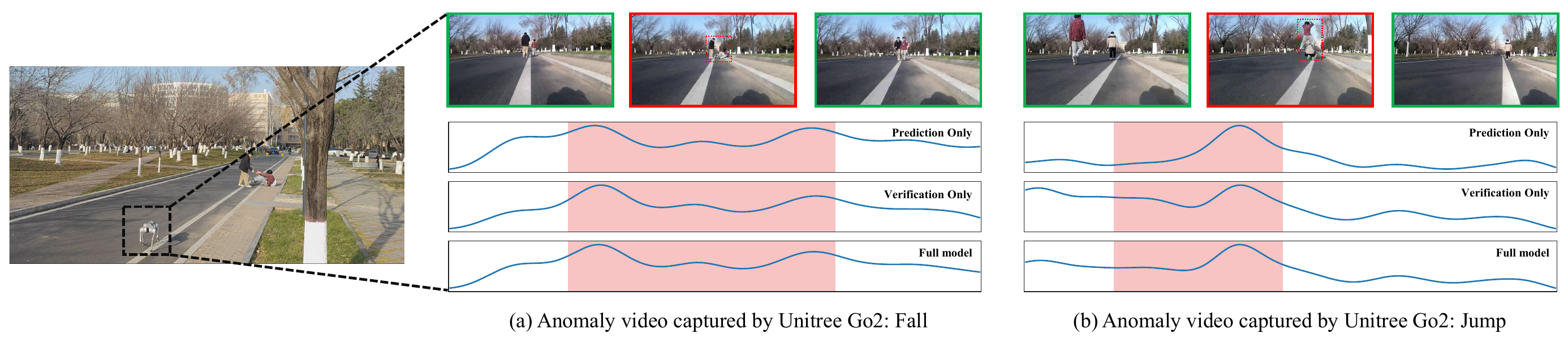}
    \caption{Preliminary evaluation on a real-world Unitree Go2 quadruped robot platform. Left: the robot during campus patrol. Right: two first-person anomaly sequences—(a) Fall and (b) Jump—showing sample frames (green borders: normal, red borders: anomalous) and anomaly score curves from Prediction Only, Verification Only, and the proposed full INTERACT solution. Pink shaded regions indicate ground-truth anomalous intervals.}
    \label{fig:robot_example}
\end{figure*}

Table~\ref{tab:ablation} evaluates each component on ShanghaiTech. Variant A1 evaluates an entity-independent setting in which information exchange across entities is disabled during representation construction and entity prediction. It yields $79.18\%$ AUC, a drop of $5.32$ points---the largest degradation among all ablations.
Variant A2 removes the MIRA verification branch and retains only the Stage~1 prediction scores, achieving $80.93\%$ AUC, a $3.57$-point drop. Variant A3 removes the Stage~1 prediction scores (both entity-level and global-level) and retains only the MIRA verification scores, achieving $82.93\%$ AUC, a $1.57$-point drop. Variant A4 removes the semantic consistency loss during Stage~2 training while retaining the corresponding reconstruction score during inference, yielding $83.43\%$ AUC, a $1.07$-point drop.
INTERACT consistently outperforms both single-branch configurations, confirming the complementary nature of prediction and verification pathways.

\textbf{Qualitative analysis.}
In Fig.~\ref{fig:entity_relations}, the anomalous object O1 in both scenes contains predominantly blue regions. Occluding these regions lowers the Stage~1 anomaly score, suggesting that their appearance and motion cues contribute to the original anomaly response.

\subsection{Computational Efficiency}

\begin{table}[!t]
\centering
\caption{Computational efficiency of INTERACT.}
\label{tab:efficiency}
\begin{tabular}{cccc}
\toprule
Entities & Params (M) & GFLOPs/clip & FPS \\
\midrule
1 & 57.18 & 41.84 & 8.61 \\
3 & 57.18 & 61.84 & 5.75 \\
5 & 57.18 & 81.73 & 5.30 \\
\bottomrule
\end{tabular}
\par\vspace{5pt}
\parbox{\columnwidth}{\footnotesize Synthetic 16-frame inputs on an RTX 4090; FP32; batch size one. FPS counts target-frame scores before temporal smoothing and excludes input preparation, transfers, YOLOv8, and RAFT.}
\end{table}

Table~\ref{tab:efficiency} shows increasing computation across entity configurations with a fixed parameter count. Stage~1 accounts for approximately $90\%$ of full-model latency in the three-entity setting. Adding MIRA increases latency by $17.09$~ms ($10.9\%$) and introduces $4.96$M parameters, with POAMI reused for semantic verification. The corresponding ShanghaiTech ablation in Table~\ref{tab:ablation} improves AUC from $80.93\%$ to $84.50\%$, supporting the verification branch's contribution relative to its incremental runtime cost. GFLOPs are estimated from the principal multiply-accumulate operations, counting two operations per MAC.

\subsection{Preliminary Evaluation on a Quadruped Robot}

To examine the potential for extending INTERACT to mobile embodied platforms, we conducted a preliminary evaluation on first-person videos captured by a quadruped robot. 
The videos were captured using the built-in camera of a Unitree Go2 quadruped robot (Fig.~\ref{fig:robot_example}). 
The videos were recorded at 720p resolution and $30$ frames per second. The Fall video contains frames 0--200, with frames 46--146 inclusive manually annotated as anomalous. The Jump video contains frames 0--207, with frames 36--101 inclusive manually annotated as anomalous. Without retraining or fine-tuning, the ShanghaiTech-trained INTERACT model was directly applied using the same YOLOv8 detector, RAFT optical-flow extractor, score fusion, and Gaussian smoothing procedure as in the benchmark experiments. AUC is computed separately within each video. INTERACT achieves $74.0\%$ AUC on Fall and $84.9\%$ on Jump.

These results provide a preliminary indication that the learned representations remain informative under a change from fixed surveillance views to a mobile first-person viewpoint.

\section{Conclusion}

We presented INTERACT (Interaction-Centric Network for Temporal Entity-Relation Analysis and Consistency Testing) for unsupervised video anomaly detection in surveillance and robotic monitoring. By jointly modeling entity-level appearance and motion cues with dependencies across entities, INTERACT assesses normality through interaction predictability, conditional motion reconstruction, and semantic consistency.

Experiments on ShanghaiTech, Avenue, and NWPU Campus\_sub evaluate INTERACT across diverse anomaly types, achieving 84.5\% frame-level AUC on ShanghaiTech. Disabling cross-entity information exchange reduces AUC by 5.32 percentage points, supporting the contribution of relational modeling. A preliminary evaluation on two Unitree Go2 sequences further suggests potential for transfer to mobile first-person views without retraining, although broader evaluation is needed to establish generalization.

Future work will investigate longer temporal context, more diverse mobile-platform scenarios, and scene-aware extensions for robotic patrol applications.

\FloatBarrier
\bibliographystyle{IEEEtran}
\bibliography{bib}

\end{document}